# Data management in systems biology I – Overview and bibliography

Gerhard Mayer, University of Stuttgart, Institute of Biochemical Engineering (IBVT), Allmandring 31, D-70569 Stuttgart

**Abstract**
Large systems biology projects can encompass several workgroups often located in different countries. An overview about existing data standards in systems biology and the management, storage, exchange and integration of the generated data in large distributed research projects is given, the pros and cons of the different approaches are illustrated from a practical point of view, the existing software – open source as well as commercial - and the relevant literature is extensively overviewed, so that the reader should be enabled to decide which data management approach is the best suited for his special needs. An emphasis is laid on the use of workflow systems and of TAB-based formats. The data in this format can be viewed and edited easily using spreadsheet programs which are familiar to the working experimental biologists. The use of workflows for the standardized access to data in either own or publicly available databanks and the standardization of operation procedures is presented. The use of ontologies and semantic web technologies for data management will be discussed in a further paper.

## INTRODUCTION

The large amount of data produced by biological research projects grows at a fast rate. The 2009 edition of the annual *Nucleic Acids Research* database issue mentions 1170 databases [1]; alone 293 databases about biological pathways are currently listed on the pathguide website [2]. Even a moderately sized pharmaceutical company could have around 1000 in-house databases [3]. With the exponential growth of the produced experimental data in all areas of modern biosciences, data management techniques are becoming more and more a necessity to enable efficient and reproducible usage of these data. This encompasses the acquisition, cleansing, administration, curation, storage, retrieval and analysis of these data in a reasonable and easy to handle way for the working biologist [4-14]. The incorporation of metadata – data about data - is of great importance. In systems biology the integration of data from heterogeneous data sources is a further great challenge. Therefore Lincoln Stein formulated the need to build up a bioinformatics nation in his paper [15] where he proposed to use web services as basic technology for the access to biological data bases. It's clear that there is an urgent need for defining standard data formats for storage, retrieval and exchange of experimental data to allow integrated data analysis procedures and methods to extract as most as possible information from the existing and coming amounts of data and to allow the generation of new information from them.

The ever increasing importance of data management approaches is reflected in several special issues devoted to this topic [16-21], in the efforts to implement educational studies dedicated especially to biological databases and curation tasks [22,23], the foundation of a new journal about biological databases [24], the foundation of the ISB (International Society for Biocuration) and special conferences like DILS (Data Integration in the Life Sciences) [25]. It can be expected that in future with the rise of personalized medicine data management solutions are becoming extremely important also in the management of clinical data to support clinical decision support systems [26].

## DEFINITION OF DATA MANAGEMENT

The Data Management Association (DAMA) [27] defines data management as "… the development, execution and supervision of plans, policies, programs and practices that control, protect, deliver and enhance the value of data and information assets". Typical topics in data as well as in meta-data management are the database architecture and administration, data modelling, data mining, data analysis and integration, data quality assurance and data security.

## DATABASES (DB's)

The typical architecture of a DB is based on the classical ANSI / SPARC three-layer model [28] consisting of a physical (mapping to the hardware), a logical (definition of the conceptual scheme) and an external layer (user interface and the APIs). The main data models used in biology are the simple flat files, more flexible XML-based text formats, the relational (e.g. Oracle, DB2, SQLServer, mySQL [29], PostgreSQL [30]) and the object-oriented model (e.g. db4o [31]). The relational model [32] is based on the relational algebra [33] and represents the data in table form, where the rows correspond to the data records and the columns to the data fields of such a record. For querying, the Structured Query

Language (SQL), consisting of Data Definition, Data Query, Data Manipulation and Data Control Language is used. For modelling the data, methods like Entity-Relationship (ER) Modelling and the theory of the relational normal forms are applied [34]. Specifics of biological DB design are overviewed in [35, 36]. A DB design method tailored to systems biology needs is described in [37]. These topics and other data models are covered in detail by standard textbooks about DB management systems [38, 39].

Typical tools supporting the modelling process are e.g. TOAD [40], DB Designer [41], ERwin Data Modeller [42], PowerDesigner [43], ER/Studio [44] and graphical frontends like Altova databasespy [45], razorSQL [46], Oracle SQL Developer Data Modeler [47], phpMyAdmin [48] for the MySQL [49] database and phpPgAdmin [50] for the PostgreSQL [51] database.

An important topic using DB systems is the data cleansing process [52] to detect errors and inconsistencies in order to ensure data quality both in terms of reliability and validity. This can be achieved for instance by integrity constraints, duplicate detection and plausibility checks and is still is an area of active research [53].

## PROBLEMS IN BIOLOGICAL DATA MANAGEMENT

The greatest problems in data management for biological and preclinical research are the enormous heterogeneity and complexity of the data, the huge volume and context dependency of the data, the data provenance [220], the nosiness of the data generated by modern high-throughput methods, the lack of well-established data standards and globally unique identifiers, which would allow easy mapping and data integration [360]. Even in the area of clinical data management, where the data must be conformant to the standards set by the FDA, there is a lack of industry-wide standards for EDC (Electronic Data Collection) and CTMS (Clinical Trials Management Systems) [54,55] systems like e.g. edc$^+$ from DSG [56]. In the preclinical area and especially the fields of metabolomics and systems biology the situation is even worse. Here one often faces vendor-specific formats and for the efficient processing of graph-based data - used for the representation of networks - one needs new data structures for the mapping of these data to the conventional relational database model [57]. The enormous complexity and variety of biological data sources poses many open data representation questions in attempts for data integration projects [58]. An overview for data models representing pathway data is given by [59].

## DATA STANDARDS

For publication most journals nowadays require that the underlying data are made available to the public in a standard compliant format [60]. The same is true for getting sponsorship from funding agencies. In the past years a lot of minimal requirements describing such data were defined, e.g. MIAME [61] (Minimum Information About a Microarray Experiment) for microarray experiments, MIAPE (Minimum Information About a Proteomics Experiment) [62,63] for proteomics data, MIRIAM (Minimum Information Requested In the Annotation of biochemical Models) [64] for biochemical kinetic models, MIGS (Minimum Information about a Genome Sequence) [65] for the description of gene sequences of the GSC (Genomics Standards Consortium) and other new sequence formats [66,67], and MIASE (Minimum Information About a Simulation Experiment) [68] for the description of simulation experiments. In the metabolomics domain the standardization process isn't as much progressed – a comprehensive coming MIAMET (Minimum Information about a Metabolomics experiment) standard is urgently needed [69-74]. Until now only some metabolomics aspects are comprised by the existing ArMet [75, 76] (Architecture for Metabolomics), SMRS [77] (Standard Metabolic Reporting Structure) and CIMR [78] (Core Information for Metabolomics Reporting) specifications. For mass spectroscopy data [79] standards like mzXML [80] and mzData [81] are used besides of NetCDF (Network Common Data Format) [82]. These formats and data are supported and integrated by the MeltDB metabolomics platform [83]. MASPECTRAS [84] and myProMS [85] are systems for use in LC_MS/MS data management for proteomics. Special data management solutions exist also for flow cytometry [86] and the management of image data [417-424].

In the last time the integration of such data standards was requested [87, 88] and led to the definition of the MIBBI [89] (Minimum Information for Biological and Biomedical Investigations) project. The MIBBI web site currently has 29 such minimum information projects registered. Further minimum information projects currently under consideration, are listed for instance on a Nature Biotechnology web site [90]. The goals of these projects are more transparency by allowing the reproduction of experiments in complementation to the good research reporting guidelines for the publications itself [91], the secondary use of data, more easily data exchange and data integration and the vision of easier searchability of such data in a future semantic web by the use of ontologies.

In general the following 3 types of data have to be standardized: Minimum information checklists, covered by the various MIBBI standards, which comprise data and metadata describing an experiment, file formats, which define the syntax of the data and ontologies for the description of the semantic level of the data. For metadata a basic standard was defined by the DCMI (Dublin Core Metadata Initiative) [92]. XMDR (eXtend3ed MetaData Registry) [93] is an implementation of the ISO standard for a MDR (MetaData Repository) as defined by ISO/IEC 11179 [94]. The various data file formats are mostly XML-based markup languages, which are defined by the various standardizing groups, e.g. MGED, (Microarray Gene Expression Data Society) for microarray data, PSI (Proteomics Standards Initiative) for proteomics data, MSI (Metabolomics Standards Initiative) [73] for metabolome data and phyloXML [95] for the representation of evolutionary (phylogenetic) trees. The minimum information specifications define which kinds of data should be documented, but normally do not describe the underlying data formats, e.g. the MIRIAM specification for biochemical models do not specify if one should code these data in SBML [96,97] (Systems Biology Markup Language), CSML (Cell System Markup Language) [98] or CellML [99,100] format. These XML-based formats are designed for processing by computers, i.e. SBML can be manipulated by the libSBML library [101]. An example of such manipulations is the merging of SBML models [102]. For human perception there is a need for visualization tools - for instance based on XSLT (eXtensible Stylesheet Language Transformations) - for such XML-based formats. An example is the tool SBML2LATEX [103] for the visualization of SBML model files. Another possibility is the presentation using a TAB-based format, which can be easily processed by simple spreadsheet programs and is very familiar to the working biologist as described in one of the following sections of this paper. ISML (In Silico Markup Language) [104] is a format which can be converted into CellML, SBML, LaTeX and also directly into parallelized C++ programs (MPI-C++), so that the models can be directly and efficiently run for simulations. The use of a XML schema for the exchange of genome mapping data is proposed in [105].

The future challenge ahead would be the integration of the mentioned standards into global data models of cellular behaviour [106]. One step in this direction are comprehensive standards like for instance BioPAX [107], which describes data at various levels of biological pathways like metabolic pathways, molecular interaction networks, signal transduction networks and genetic regulatory networks. A future fifth level of BioPAX is planned to represent data about cell-level interactions and the integration of environmental effects. As shown in [108] with increasing level the BioPAX standard subsumes more and more of some other standards [109] like for instance the PSI-MI standard for molecular interaction of the Proteomics Standards Initiative [110]. PaxTools is a software library supporting the access to and manipulation of BioPAX models and can be downloaded from sourceforge [111]. An extension of BioPAX called SBPAX (Systems Biology Pathway Exchange) is currently under development as part of the Virtual Cell [112] modelling and simulation environment, allow taking full advantage of existing pathway data in kinetic modelling [113]. Some more specialised standards are SBRML (System Biology Results Markup Language) [114], SED-ML (Simulation Experiment Description Markup Language) [115], which implements the MIASE guidelines, Strenda (Standards for reporting Enzymology data) [116], BPR (Batch Process Record) and BMR (Batch Manufacturing Record) for production processes [117] as part of the GMP (Good Manufacturing Practices) requirements of the FDA or the Metabolic Flux Analysis Markup Language MFAML for the representation and exchange of metabolic flux models [118].

Whereas the former mentioned data standards aim at the biological practitioners there are some further comprehensive standardization efforts under development, like the IEEE–1953 BSC (Bioinformatics Standards Committee) [119] standard which defines a reference system on bioinformatics data structures aiming at the implementers, i.e. software engineers who want to provide software tools implementing the various standards.

Other standards describe not only the data but also the protocols and procedures of laboratory techniques [120], e.g. for microarray [121-123] and tissue microarray [124,125] data analysis and management. There are several websites for sharing such protocols [126-129].

Detailed overviews about systems biology standards are given in [130-137]. The process of defining standards for experimental protocols is exemplified in [138]. A new trend is the use of DSL's (Domain Specific Languages) for the rapid implementation of newly evolving standards as recommended in [139]. DSL's are languages which allow extending their syntax to support the embedding of custom language statements adapted to one's problem domain. Typical programming languages suitable for the development of such DSL's are the functional languages F# [140] and the scripting languages Boo [141] and Groovy [142], which support the so called language-oriented programming paradigm [143] by incorporating built-in support for scanners, parsers and AST's (Abstract Syntax Trees), which are representations of the syntactic structure of programs.

**DATA MANAGEMENT**

Following an overview over the general possibilities of data management systems in systems biology is given.

*1. Spreadsheet-based approaches*: This is the simplest approach based on using spreadsheet programs, which are familiar to most of the people working in a laboratory. This includes the use of template spread sheets like the TAB-based formats MAGE-TAB and ISA-TAB, which are described in a separate section below.

*2. Web-based document-sharing tools*: These are either the open-source wikis [144,145], semantic wikis [146,147], which combine wikis with semantic web technologies or groupware programs like eGroupware [148], PHProjekt [149], BaseCamp [150], Alfresco [151], DaMaSys (Data Management Systems biology) [152] or BSCW (Be Smart Cooperate Worldwide) [153]. These allow distributed project teams to exchange and access the data via simple web browsers. Of main importance is the possibility to define groups and to give the users and groups access rights, so that the privacy of the data is ensured until the results based on the data are published. Afterwards the data can be copied into a public group, so that the requirements of the journals for data transparency can be easily fulfilled. The strengths of such tools are the exchange of SOP's (Standard Operating Procedures) and experimental protocols and for project management tasks. Besides document-sharing these tools also offer project management functionalities. Often these document management tools also offer versioning of the documents. Because they are no real databases they have the disadvantage that measures to ensure data consistency between the data submitted by different users [154] are missing. Therefore they are not suitable for the management of experimental data. It should be clear that because of data security reasons one shouldn't use online services like Windows Live, Google Apps and related services for sharing documents.

*3. Laboratory Information Management Systems (LIMS) / Laboratory Information Systems (LIS) / Electronic Lab Notebooks (ELN)*: Whereas the focus of LIMS's and ELN's is on biological and biochemical laboratories , **LIS's** are systems designed for medical and hospital laboratories. *Cap Today* gives annually an overview about existing LIS systems [155].

**ELN's** [156,157] are a replacement of traditional laboratory notebooks. They have the benefits to be searchable, to enforce an elementary standardization of record keeping and allow the use of predesigned user-controlled templates [158]. The ultimate vision of ELN's is the 'paperless web'. A key requirement for ELN's is the ease of use and the integration with the existing workflows [159]. Further challenges introducing ELN's in a laboratory environment are described in [160]. A widely used ELN is the CERF (Collaborative Electronic Research Framework) notebook from Rescentris Ltd. [161], which is based on the BSML (Bioinformatics Sequence Markup Language) and is compliant with 21 CFR Part 11 (Code of Federal Regulations) [162] of the FDA, so that it can be used in GxP environments, which is of prime importance for the use in the pharmaceutical industry. Further standards for ELN's are defined by the CENSA (Collaborative Electronic Notebook Systems Association) [163].

**LIMS** systems allow the organization of all generated data inside a laboratory. There are a lot of commercial and open source systems available, see the limsource [164], limsfinder [165] or Laboratory Informatics Guide [166] websites for an overview of existing LIMS systems. Open source examples are Sesame [167-169] and SetupX [170] for metabolomics and Bika [171] for a broader application spectrum. Besides project management (administration of users and user groups) such LIMS systems allow the management of samples, instruments, standards, plate management, work flow automation and for commercial environments they also support functions like invoicing. There is a great variety of LIMS's; a lot of them are very specialized, like Xtrack [172] and HalX [173-175], which focus on crystallography and structural genomics, so that by far not all LIMS systems are suitable in a systems biology environment, where the requirement is the integration of various data from the transcriptomics, proteomics, metabolomics and modelling domain. Bika [171] is a combined LIMS, workflow and content management system, which is available for different branches like chemistry and biochemistry, beverages, health laboratories, geology, petro and mining. The disadvantage of Bika is its complexity, so that one needs access to bioinformatics support if one wants to configure it to its own requirements. A Perl toolkit for development of in-house LIMS systems is also available [176].

*4. Web-services based workflow systems*: One application of the grid computing concept [177-179] are scientific workflow / pipelining systems like

Triana [180], Kepler [181], caGrid [182], KNIME (Konstanz Information Miner) [183]and others [184], which are an essential part of the eScience vision [185] allowing to perform experiments in silico [186,187]. There are specialized workflow systems mirroring the research pipeline processes of e.g. structural biology [188] or flow cytometry [189] as well as more general bioinformatics workflow systems like Taverna, Wildfire [190], BioWMS [191], O2I (Oncology over Internet) [192], BioWBI [193] from IBM and SOMA2 [194]. Complete overviews about workflow frameworks in the life sciences are given in [195-197].

The most prominent and for systems biology best suited system is Taverna [198-200], which allows the access to over 3500 data sources on the internet and the construction of workflows for analysing and integrating these data. With Taverna repetitive tasks can be standardized by the provision of standard workflows, which can be seen as SOP's (Standard Operating Procedures) for data analysis. The myExperiment website [201] can be used as a resource for the exchange of such workflows; an alternative to myExperiment is the Workflow Enactment Portal BioWep [202]. Taverna itself offers the possibility to access web services [203] described by WSDL (Web Services Description Language) files as well as the access via the SOAP (Simple Open Access Protocol) protocol [204]. In addition Taverna supports Soaplab [205], a framework for access via web services to command-line bioinformatics programs, like for instance the EMBOSS (European Molecular Biology Open Source Software) [206] suite of programs. Also the BioMOBY [207-209] system, which is based on ontology-based messaging and which allows the automatic discovery of appropriate data and / or analytical service providers [210,211] can be accessed through Taverna. In addition Taverna allows the nesting of workflows, so that these can be developed in a modularized style. Furthermore data bases can be accessed via BioMart [212], an interface which allows uniform access to databases via a web interface. Local workers can be embedded into Taverna workflows by using BeanShell [213] scripts or Java API consumers. Last but not least an interface for accessing the statistical R project [214] and the Bioconductor [215] statistical analysis tools is part of Taverna [216]. An interface for directly accessing Matlab [217] resp. its open source alternative Gnu Octave [218] is in preparation. Taverna furthermore allows querying XML databases [219], interfacing with computationally demanding analyses running on a grid environment [220], the incorporation of manual interactions in workflows [221] and the manipulation of SBML models [222]. An important issue is the tracking of the provenance of the generated intermediate and final data [223-225]. Because biological facts are often changing, for instance due to new findings or because experimental data are often imprecise, context-dependent and incomplete, it is necessary to keep track of the evolution of the generated knowledge in order to ensure the consistency of the knowledge stored in databases and to determine the trust one can place on the derived results, which is the task of data provenance [226-228].

Newer workflow developments are Bio-STEER, which uses semantic web service technology, i.e. makes use of ontologies in order to try to alleviate the process of workflow construction for the working life scientists [229] and e-BioFlow [230]. Another approach is a rule-based one to support the designing and creating of new workflows [231]. MicroGen [232] is an example of a MIAME-compliant workflow system for microarray research. Such workflow-based systems are of increasing interest for the pharmaceutical industry for automating the drug discovery [233-235] and drug design [236] processes.

The available web services are listed in the annually web server issue of the *Nuclei Acids Research* journal and available on the Bioinformatics Links Directory [237]. The web services available from EBI (European Bioinformatics Institute) are described in [238]. The BioCatalogue website is a curated catalogue of available life science web services worldwide [239,240].

In summary workflow-based systems mainly concentrate on data analysis tasks. Using them for data storage and retrieval applications can be easily reached by combining them via BioMart [212] with databases.

But one main task remains to be done to bring Taverna to its full potential: Today most newly developed data analysis programs described in publications are not available with a web service interface so that they cannot be integrated into Taverna workflows. Therefore the journals should require as a prerequisite for publication not only that newly developed software is made freely available as source code or for download but also that it has a web service interface allowing easy integration into Taverna or other workflow systems. One needs not only standard repositories for data and models but also for software programs acting on them. For instance imagine you can easily construct a Taverna workflow for analysing microarray data with the option to choose from the wealth of analysis methods already developed and published. Today most PhD students either work as experimentalist or they develop an algorithm implementing a new and clever analysis method. The experimentalists store their results in databases like GEO or ArrayExpress.

The software developers mostly publish a paper and maybe put their software onto their own website for download. It would be much better if the developers integrate a web service interface into their software and then put it together with a concise documentation in a public directory where everyone can access the functionality provided by the software and can use it by integrating into their workflow system. Only if researchers have free and easy access to both data and analysis software they are animated to build Taverna workflows which in turn can be made available at sites like myExperiment. Such an approach would promote other researchers to make use of the data graveyards we have today and would ensure that in future a lot of new knowledge and insight is gained only by analysing the already existing data with already existing software. This is exactly what's behind the SOA idea: simply reuse of software pieces and building them together to larger analysis pipelines by only knowing how to access the software via a clearly defined web interface independently from the programming language which was used to implement the software. This would be a further big advantage of such a repository: if a developer wants to integrate a piece of software into his own software he easily can do this simply by using his preferred programming language which only must support the use of web services via WSDL files.

## CHOOSING A DATA MANAGEMENT SYSTEM

If one has to decide which data management system to use for a planned project, one has in principle the following three possibilities:

*1. Developing a proprietary solution*: This is the most expensive solution and requires that one has the needed manpower for development and maintenance at hand, but on the other side it gives one the security getting a system that exactly fulfils ones requirements. Such a proprietary solution is only recommended if one needs functionality not offered by open source or commercial standard software. An example is the Systems Biology Institute at Seattle which developed SBEAMS [241] (Systems Biology Experiment Analysis Management System), an integrated system that combines a relational DBMS backend with a web front-end for managing and processing their microarray and proteomics data. Another such an example is the AMEN [242] suite of tools. Also a lot of pharma companies implemented their own solutions, as e.g. described in [243] or the PEKE (Pfizer Environment for Knowledge Engineering) of Pfizer Inc. [244].

*2. Use of open source software*:
As an alternative to develop individual software solutions, one can use freely available software. Both SBEAMS and AMEN are such packages, which can be used freely and possibly customized to ones own requirements. Another example in the transcriptomics area are the ArrayTrack$^{TM}$ software [245,246] developed and used by the FDA (Food and Drug Administration) for genomic data submissions and the BRB-Array Tools [247] of the NCI (National Cancer Institute). An overview about microarray data management is given by [248]. In academic research the use of open source is already widespread and because the NIH (National Institutes of Health) mandates open access for results of NIH-funded research it is expected that open source data and software will be of increasing importance in future even in pharmaceutical drug discovery programs [249,250], paving the way to open source science by public-private partnerships [251].
An open source data integration software is Talend Open Studio [252], which allows one to integrate data from diverse sources based on an ETL (Extract, Transform, Load) approach so that it supports the building of data warehouses. This software is freely usable, but also very complex, so that services for it are marketed by Talend.

*3. Use of commercial standard software*: Another possibility is the use of standard software for data management tasks. An example is the Genedata suite [253] consisting of Expressionist, Phylosopher and Screener. These are integrative software packages with a multitude of features integrating transcriptomics, proteomic, metabolomic and phenotypic data and widespread use in biotechnology and pharmaceutical industry for data management and biomarker discovery. Expressionist allows the management, analysis and visualization of experimental microarray, protein 2D gel, mass spectroscopic and chromatography data. For the use of academic research projects the Genedata software is maybe less suited because of its high price. Furthermore if used over internet connections the performance of the suite is moderate, even if it performs very well in an intranet environment because Genedata uses the slow-going WebStart technology, which allows using the familiar Swing GUI of Java applications, but brings a considerable performance lost, even if used via broadband internet connections. Other widespread used commercial products are GeneLogics [254], GeneBio [255], Accelrys Pipeline Pilot [256] for data integration, analysis and reporting tasks and InforSense [257]. Of course there is a wealth of other commercial products [2], often with a more specialized focus like text mining or pathway analysis, e.g. the Ingenuity Pathway Analysis Suite [258]. Data management solutions for protein antibody research are reviewed in [259]. In addition IT service and database companies offer special solutions for the biosciences, e.g. BioWBI [193] and the WebSphere Information Integrator [260] from IBM, the Oracle Life Sciences Platform [261] based on the Oracle database 11g or the Data Integration Suite [262] from DataDirect Technologies for data integration tasks. OpenLink has made its Virtuoso [263] data integration and data management solution open source. IBM offers DataStage as data integration and management platform [264] and Microsoft is planning to market Amalga LS [265] as a R&D and trial data management platform [266].

A problem of such commercial packages for the use in innovative distributed research projects like for instance HepatoSys [267] or SysMo [268] is that newer methods and their associated formats, which haven't found yet the way into industrial practice, are often not yet supported. In addition users in such distributed academic research projects often insist to use the data analysis tools they are familiar with, so that it's not easy to bring them to use such a software package in its full content. Therefore one must ask if it's not better to use in such cases free open source packages like data and document sharing tools, which only allow the exchange of data. Then for analysis purposes they are free to either use other workflow-based tools like Taverna or to use their own commercial products, often provided by the device manufacturers, with which they are familiar with. If one decides to use a commercial product one should choose a potent partner to make sure that steady support and updates are given – remember that erstwhile celebrated companies like Lion Bioscience and others quickly disappeared from the market. On the other side also open source projects often loose funding support after some years [269].

## DATA INTEGRATION

With increasing speed of high throughput experiments data integration and data mining become more and more important [270]. According to [271,272] there exist 4 principal strategic ways for data integration [273]:

*1. Link integration*: This is the simplest way of data integration, based of interlinking to build up cross-referencing systems like e.g. SRS (Sequence Retrieval System) [274] and Entrez [275].

*2. View integration (federated databases)*: An uniform view on several DB's is constructed which appears to the user as one database. The user queries are translated into cross-database query language. Examples are TAMBIS [276] and Kleisli [277]. The disadvantage is that the overall performance is limited by the slowest source database.

*3. Data warehousing*: Here the source databases are regularly scanned and loaded into a central database by an ETL (Extract, Transform, Load) procedure. This requires a predetermined unified data model and requires regular updates, so that the maintenance costs are high because one must adopt the loading procedures if a source database changes its data model [278]. Examples are the Atlas [279] and the EnsMart [280] systems. For data warehouse it is of importance to keep track of where each unit of data came from (provenance information). An open source toolkit for building data warehouses is described in [281]. Data warehousing for microarray data is described in [282].

*4. Web services, Service oriented architecture (SOA, WOA)*: Web services are programmatic web interfaces, allowing integration of data by e.g. workflow programs. An example is the Bioverse API (Application Programming Interface) [283]. The web services are based on protocols like WSDL, SOAP and UDDI (Universal Description, Discovery and Integration). In the Web 2.0 field often simpler protocols like AJAX (Asynchronous Java and XML) [284] and REST (Representational State Transfer) [285] or XML-RPC (Remote Procedure Call) [286], which can be seen as a lightweight version of SOAP, are used together with simple data interchange formats like JSON (JavaScript Object Notation) [287]. Unfortunately the web service interfaces are often poorly documented which aggravates the widespread use of integration pipelines based on workflow systems like Taverna [198-200] by biologists. Web services should alleviate the users to find the right programs to use, the parameters to tune these programs and help them to manage the input / output formats as expected by these programs [288]. In principle one can find all the relevant information in the WSDL files, but without good tool support this can be cumbersome, especially for people not well trained in the use of such WSDL files. Provided that this documentation resp. tool support lack could be resolved, that repositories for software analysis programs are provided and that such workflow based computational tasks become part of biologists curricula, systems like Taverna have the potential to become something as a killer application during the next decade in the bioscience field.

Another approach tried by the Seahawk system [289] is to build workflow systems, which can be used by biologists without the assistance of programmers / workflow developers. This too requires a standard mechanism, i.e. a repository by which analysis software is made available to the research community in a standardized way. Such a "Minimum Information about a Software Program" (MISP) and a central repository for depositing such MISP concordant programs we miss painfully today. Such a MISP specification should require that the software must have a web service interface described by WSDL files and an accompanying documentation for workflow developers. Furthermore beside administrative information about the developer, the used programming language, the version, the memory and runtime complexity, the platform on which to run the software, the possibility to execute the software in a grid or cloud environment and on which environment (e.g. GoogleApps [290,291], E2C [292], Windows Azure [293], …), the intended use of the software and so on it should be required that the supported input and output formats are clearly specified, so that integration into workflow analysis pipelines is eased. Therefore in my opinion such a MISP specification and a dedicated repository for analysis programs is a prerequisite for making systems like Taverna the killer application for the life science research of tomorrow. One proposal for making web services more usable for end-users is PoSR (Potsdam Services Registry) [294]. In addition an ontology for the clear classification of the exact tasks that such software fulfils would be very helpful by offering the possibility to enrich the repository with a semantic search machine for allowing users or in future even software agents of the proposed biological analysis software repository to easily find the suitable program they need. The annotation of the software programs with such ontological terms should also be part of the proposed MISP specification. A future vision and a continuing development of software agents using such repositories would be an in silico robot scientist [295], autonomously applying software analysing the data stored in the biological databases. **SOA** is a software architecture based on web services [296]. An example for such a SOA – based system is caCORE [297], a part of the cancer Biomedical Informatics Grid (caBIG) of the NCI (National Cancer Institute). A subset of SOA is WOA, the Web-Oriented Architecture, which uses the simpler REST protocol [285] instead of the web services, which are based on the more complicated SOAP and RPC (Remote Procedure Call) protocols for communications via the web. In short one can define **WOA** = SOA + WWW + REST.

Besides these 4 basic data integration approaches there are some other approaches:

*5. Loosely coupled systems*: For instance the Gaggle system [298] uses a minimalistic approach to integrate diverse databases and software by mapping to only four basic data types (names, matrices, networks and associative arrays) and by using only modest adaptations of existing web resources. Java RMI (Remote Method Invocation) [299] and WebStart [300] are used as the underlying integration technology. A similar approach is pursuited by the GENE-EYE system with the goal to conserve flexibility for the easy extension of accruing data sources when biological progress requires the integration of new data types [301]. Due to the steadily evolving experimental techniques in systems biology readily adaptable data management systems are required and this adaptability is the great advantage of such loosely coupled [302] and lightweight integration systems [303]. An alternative for representation of highly complex, heterogeneous and rapidly evolving data schemas is the EAV/CR (Entity-Attribute-Value with Classes and Relationships) [304] representation because of its high flexibility regarding to data schema redesign.

*6. Visualization*: Standardization efforts are required for the visualization of data, e.g. the layout extensions [305,306] for SBML models and the SBGN (Systems Biology Graphical Notation) [307,308] as standard for the presentation of biochemical models. Such visualization approaches can also be used for data integrative tasks. Examples are the ONDEX suite [309] by which experimental data from diverse experimental data sets can be linked and visualized and analyzed in an integrated manner based on graph analysis techniques. An enhancement is the ONDEX SABR project [310] currently under way. Another such a visualization tool is VisANT [311].For navigating through literature, visualization techniques can applied too, for instance iHOP (Information Hyperlinked over Proteins) [312] allows a gene-guided traversing of the literature space with regard to protein interaction networks. In systems biology the management of graph-based network data (metabolic, signalling and gene regulatory networks) is of eminent importance and special data structures are used for the management of such graph data [313,314]. An example for such a graph-based integration is the BIOZON system [315] and in [57] an extension for managing graphs in IBM DB2 database system is described. A commercial system using visualization approaches for data integration is the Ingenuity Pathway analysis software [258], which allows one for instance to map colour coded gene expression time series data on metabolic maps so that one can watch how the expression changes spread through the metabolic net with progression of time. Also some other software packages like the Genedata system and PathVisio [316] allow such analyses. Another interesting approach is GenomeMatrix [317], which allows an integrated view on heterogeneous data across several organisms. Also the visualization of gene expression data over interaction networks is often applied to show the strengths of the different interactions or gene functions [318].

Visualization tools are often used in combination with another data integration approach, e.g. SDRF2GRAPH for the visualization of ISA-Tab SDRF files [319]. Overviews about visualization tools are given in [320-322]. Open biological network visualization problems are reviewed in [323] and 3D visualization is described in [324].

*7. Agent based systems*: Agents are software acting autonomously on behalf of their users in a reactive, proactive and adaptive way, i.e. they react to their environment, they show goal-directed behaviour and they are able to learn from experience so that they improve themselves [325]. For accomplishing their work they communicate with their users and with other agents. Building up on web services and semantic web technology they can be used for information discovery and integration [326,327]. A framework for development of such agents is Jade [328], the Java Agent Development framework.

*8. Portal solutions:* Examples are the systems Biozon [329], BioNavigation [330] and BioGuideSRS [331], which builds up on the SRS system. They use either link integration together with graph structures [332] or ontologies to guide the search process through the databases. For constructing and implementing one's own user-friendly portal solution one can use the open source Liferay portal software [333] as a basis, which builds up on the Java portlet standard [334].

*9. Semantic web technologies:* Semantically annotating the data with RDF, RDFS and OWL [335,336] are becoming more and more important for reaching the goal of true data integration by allowing matching the information by their annotations. Typical example systems based on semantic technology are LinkedLifeData [337] in the systems biology and LinkHub in the proteomics field [338]. BOWiki [339] is a system supporting ontology-based annotation and integration of data. The semantic web is described in more detail in a future second paper.

The reliability of individual -omics data measurements is limited by the high noise levels due to measurement errors and the inherent stochasticity of biological systems and by the under-determination problems of these measurements [340]. The hope is that by integrative network reconstruction methods one is able to more easily cope with these problems due to the expected higher robustness of the reconstructed integrative network models. Therefore the integration of the different omics data sets (transcriptomics, proteomics, metabolomics, interactomics, RNomics, fluxomics, …) [341] is a main challenge for the future of systems biology approaches [342-345] and of comparative genomics [346].

One general problem in data integration are erroneous source data, e.g. due to annotation errors or because a lack of a standardized nomenclature. The latter case could be avoided by consequent use of ontologies. In the other cases one has to decide how to deal with conflicting values in data integration. Here one must decide either to choose the best value, to keep all values equally or to assign probabilities to the different differing values according a defined probability model.

IMG is a data management, analysis and annotation system for data resulting from microbial and metagenome sequencing projects [347-350] released by the BDMTC (Biological Data Management and Technology Center) [351]. Another challenge is the management of the mass of data generated by the newly short read technologies and future nanopore sequencing techniques [352-354].

A common requirement of all data integration methods is the conversion of different data formats. This can be reached by use of XML schemas [355-358, 105] as for example realized in the XMLPipeDB [359]. A further challenge in systems biology is to identify the common components of different data stemming from different measuring techniques. Some statistical models of such simultaneous component methods are overviewed in [360].

An outline about data integration approaches in the proteomics field is given in [361] and a recent proteomics integration system is described in [362], for functional genomics in [363,364] and for genomic medicine in [365]. Overviews about the specific challenges in biological data integration are given in [366,367].

## UNIQUE IDENTIFIERS - LSID

A problem hindering the integration of data is the lack of globally unique identifiers [368]. As long as genes and proteins have several different names it's not possible to merge clearly the data belonging to them. The OCLC (Online Community Library

Center) proposed to use PURL's (Persistent Uniform Resource Locators) [369] for this purpose. Another proposal stemming from the life science community and supported by the OMG (Object Management Group) [370] is LSID, the Life Science Identifier [371-374], which is a stable GUID (Globally Unique IDentifier). There are LSID software libraries available for the Java, .NET and Perl languages [375]. Resolving a LSID returns metadata in RDF format describing a service resource available on the web [376]. The general format of such an identifier is as follows:
<LSID>::='urn:''lsid:'<protocolID>':'<authorityNamespaceID>':'<objectID>[':'revisionID], where the parts in angle brackets are placeholders and the revision identifier is optional. An example for such a LSID is for instance
URN:LSID:ebi.ac.uk:SWISS-PROT.accession:P34355:3
Another approach is pursuited by the PORTAL-DOORS framework [377], which is a mapping of ontology resource labels to the locations of these resources. The idea is comparable to the traditional web, where a DNS (Domain Name Server) is used to resolve names to internet addresses. A similar proposal of the OKKAM project [378] is the use of an ENS (Entity Name System) which works equivalent to the DNS (Domain Name System), which resolves IP addresses, in ensuring that every entity on the web gets a globally unique identifier. A similar approach is the shared names initiative [379].

## MICROARRAY, MODEL, PATHWAY, AND IMAGE DATABASES

Databases like GEO (Gene Expression Omnibus) [380], ArrayExpress [381], CIBEX [382], the Stanford Microarray Database [383] and projects like M-Chips [384,385] allow the deposition of experimental transcriptome data accompanying a publication. For the submission of microarray data TAB-based formats like MAGE-TAB and ISA-TAB are used. An alternative is caArray of the NCI (National Cancer Institute) [376]. An overview about microarray data management is given in [387]. For the submission of pathways in SBML or BioPAX format the BioModels [388] and JWS Online [389,390] databases exist. In SBML files the sboTerm attribute can be used for annotation with BioModel qualifiers. These are the two model qualifiers (bqmodel) *is and isDescribedBy* and the ten biology qualifiers (bqbiol) *is, hasPart, isPartOf, occursIn, hasVersion, encodes, isVersionOf, isHomologTo, isDescribedBy and isEncodedBy*, which link the SBML model resp. its components to each other or to other resources like for instance literature citations. Model databases are reviewed in [391].

The SBML models from these databases can be imported, analyzed and simulated by a number of systems biology modelling and analysis suites like Bio-SPICE [392], SBW (Systems Biology Workbench) [393], PottersWheel [394] Sycamore [395] and COPASI [396,397]. A comparison of deterministic simulators is given in [398]. BioNetS [399] is a stochastic simulator for biochemical network models. The SBML models can also be converted into other formats, e.g. Beta-Binders [400] to allow process algebra-based [401-403] or rule-based temporal logic simulations [404]. An overview about simulation methods for such SBML models is given by [405,406]. For merging SBML models tools like semanticSBML [407] can be used. For the analysis of BioPAX models the Cytoscape [408] plug-in BiNoM [409] can be used; ChiBE [410] is a SBGN-aware BioPAX-Editor.
Correspondingly for the analysis of the gene expression data, Bioconductor [411] can be used which provides a wealth of tools for expression analysis [412]. A comprehensive Java framework for storing, analyzing, simulating and visualization of experimental data is BioUML [413], which among others integrates access to the statistical R system [414] and Matlab. It is expected that version 1.0 of BioUML will appear in 2010.
Metabolomics databases for metabolite identification by NMR and mass spectroscopic data are the MMCD (Madison Metabolomics Consortium Database) [415] and the HMDB (Human Metabolome Database) [416].
The management of image data [417,418] becomes more and more important in systems biology, e.g. to document the subcellular location [419,420] of proteins and the dynamics of protein transport processes inside a cell by use of fluorescent microscopy images and image sequences. Typical bio-image DB's are PSLID [421], LOCATE [422] and CCDB [423]. In addition in structural biology image DB's are used for storing and analyzing 2D crystal images [424].
Other important databases for systems biology are interaction and complex databases reviewed in [425], like CORUM [426], BioGRID [427] and MiMI [428], a database which emphasizes usability issues [429]. Databases of already known pathways are KEGG [430], MetaCyc / BioCyc [431] resp. EcoCyc [432], important regulation factor databases are PRODORIC [433] and TRANSFAC [434] and useful enzyme databases are BRENDA [435] and SABIO-RK [436] containing enzyme and reaction kinetic data, which are also relevant for synthetic biology [437].
Exhaustive lists of existing biological databases are given in the annual database issue of *Nucleic Acids Research* [1,438] and on the corresponding NAR online Molecular Biology Database collection website [439], in the database of biological databases [440], in the MetaBase [441] with currently 2652 entries as well as hopefully in future by the BioRegistry project [442] which is currently built up. Another registry for semantic biological web services is SemBOWSER [443]. The Bio Netbook website [444] of the Pasteur Institute is a directory of biology related web pages and currently links to 785 database sites. A list of pathway databases can be found on the pathguide [2] or on the pathwaycommons [445] website. The EBI databases can be found on [446] and the NCBI DB's on [447].

## TAB-BASED FORMATS

As already mentioned in the last section TAB-based formats are gaining more and more importance. The first such format was SDTM (Study Data Tabulation Model) of the CDISC (Clinical Data Interchange Standards Consortium) [448] for the voluntary electronic submission of clinical study data to the FDA. Then MAGE-TAB (MicroArray Gene Expression) for microarray data submissions and recently ISA-TAB [449,450] (Investigation – Study - Assay) which supports also other data types than microarrays were introduced. The advantage is that these formats are much easier than the more complex XML-based formats, e.g. MAGE-ML, that they are easily human-readable, that they can be processed by simple spreadsheet programs which are familiar to the biologists and that the use of templates is possible. The conversion to XML-based or other DB-internal formats is then done by automatic programs. The other way around, such TAB-based formats can be used for the presentation of XML-based formats to the user by the use of XSLT or other transformations [451] like STX (Streaming Transformations for XML) [452]. As an example the implementation of the MIAME requirements by MAGE-TAB [453] is demonstrated in the following: MIAME contains information about:
- the design of the experiment as a whole e.g. experimental factors as time, dose, compound or disease state
- the array design, i.e. the assignment of genes to spot positions, as well as dye swap design, the number of repetitions, …
- sample descriptions: the samples used, the extract preparation and the labelling procedures
- hybridization description: the hybridization procedure and the parameters used
- The raw and processed data (quantification, normalization, …)

These by MIAME required data are directly reflected in the different MAGE-TAB files:
- Investigational Description Format (IDF) file containing the name of the experimenter, a description of the experiment, bibliographic references, …
- Array Design Format (ADF) file containing the assignment of sequences to positions
- Sample and Data Relationship Format (SDRF) file which describes the mapping of samples to the object data by an Investigation Design Graph (IDG)
- Raw and processed data files in ASCII or binary format: the data matrix, e.g. the Affymetrix .CEL files with rows representing genes and columns representing the time points or the experimental conditions. This is also called the FGDM (Final Gene expression Data Matrix) file.

In practice one chooses on a website hosted by the EBI (European Bioinformatics Institute) the biological design, the methodology, the used technology, the used materials and organism and then one can generate templates, which are simple spreadsheet files. These templates can then be loaded into a spreadsheet program, completed with the experimental data and then be uploaded into the corresponding microarray database.

A newer TAB-based format proposed by the RSBI (Reporting Structure for Biological Investigation) workgroup is ISA-TAB [454], a format used for the submission of experimental data via the interactive isaCreator Java tool, which supports the MIBBI specifications and the use of ontologies by integrating the OLS (Ontology Lookup Service) [455]. It is part of the BII (BioInvestigation Index) [456] infrastructure of the EBI and is named after the different sort of files used:
- *Investigation file*: contains declarative information referenced in the other files and relates several study files to an investigation
- *Study file*: contains context information for one or more assays and references to these assay files
- *Assay file*: contains information about a certain type of assay, e.g. expression or proteomics experiment including information on the used protocols and references to the data files
- *Data files*: for raw, normalized and processed data, e.g. an ADF (Array Description Format) file and the data matrix (FGDM) file according to the MAGE-Tab specification for gene expression experiments

Unlike MAGE-Tab which supports only microarray data, the assay files of ISA-TAB in addition support data from gel electrophoresis, mass spectrometry, NMR (Nuclear Magnetic Resonance) and high throughput screening experiments.
If the user has entered all information using the isaCreator program the files are zipped into an

ISArchive file, which is then transferred to the BII server. There the ISArchive file is parsed, validated and converted with isaConverter into XML-based markup-language formats and then stored in the respective databases: metabolomics data in a yet to establish MeDa database, transcript data in the ArrayExpress [457] database and proteomics data in the PRIDE [458,459] database. These databases are wrapped by BioMart [212], so that their content can be made available in a uniform way via web services to the BII, which is the interface for users retrieving the data via their web browsers. What is currently missing in the BII is a mechanism by which the submitters can lock access to their data until they make them accessible to the public. This would be a plus because then distributed research groups would be enabled to use the BII as their data management system and after they published their results based on these data they would only have to release their data to make them available to the public.

**TOWARDS OPEN SCIENCE**

It's more and more demanded by the publishers, funding organisations and the OECD [460] that all data underlying research and publications from public funded research projects should be made freely accessible. The practice today that the data are linked as supplementary material on the publisher websites has some disadvantages:
- Missing standard protocol for data sharing and terms of use for the published data [461]
- No simple to use and standardized retrieval system for such data
- Often the data made public are incomplete, so that it's hard to reproduce the results

Therefore not only the data but also the source code of the programs inclusive its documentation used to analyze the data should be made available on a central publicly accessible server in a standard semantically annotated format that includes metadata describing for instance the terms of use so that the creatorship of the depositors are preserved by clear data policy rules [462-464]. Furthermore it is requested that programs and data are permanently accessible. This can be reached for instance in a way analogous to publications where DOI's (Document Object Identifiers) [465] ensure that access to a publication is possible even if the web addresses of the servers change. By this the data would become independently citeable.

The ultimate goal of data management should be to provide easy open access to all raw and primary data to promote the eScience vision [466]. Easy retrieval is a precondition for locating and making reuse of the data. This can be reached by semantically annotation of all the data. The used semantic web techniques [467] are the topic of the second part of this review [468]. By realization of the Open Data / Open Science vision [469,470] one can increase the productivity and reliability of the scientific process. Advantages would be:
- Reuse of data for tackling additional or new questions
- Easier comparison of concurrent theories interpreting the data
- Avoidance of double work and therefore improved cost efficiency of scientific work
- Revealing of scientific misconduct [471]

The best way towards such an eScience culture would be to install publicly funded data centres in which professional biocurators annotate the data and provide the infrastructure for the storage and retrieval of all the data together with their analysis programs in a standardized fashion. Only then a long-term preservation [472] and access to the produced data can be ensured. The mentioned use of TAB-based formats at the EBI can be seen as a first step towards such a data centre for life science-related integrated data and software repository.

Recently several research groups and projects [473-478] were founded with the aim to define clear standards for such research data sharing protocols ensuring long-term access and long-term archiving for the scientific community.

What is missing today are such clear standard protocols for data set publishing, even if some proposals exist [479]. A checklist for data management plans required by grant-holders is proposed in [480].

Last it should be mentioned here that at least in the case of clinical studies negative results must be made public too in a registry- and results database as prescribed by the FDA Amendments Act.

Provided that open data becomes reality it can pave the way towards an open science culture which allows new ways of cooperation like for instanced virtual institutes, which have the potential to further accelerate and make the scientific research process more efficient.

**CONCLUSION**

One of the main challenges in data management and data integration is identity management. The mapping and matching of used terms is of prime importance. Only then can the vision of the semantic web, useful mashups, data warehouses and workflows, which often consist of identity transformation mappings, brought to its full potential. Such an identity management requires the definition of and adherence to stable, yet flexible enough, standards, ontologies and globally unique identifiers (LSID's), which allow the unique addressing of content existing in the World Wide

Web. For this the depositors of data must firmly confirm to all these standards and ontologies.

This in turn requires them to be supported by yet to develop well-established and easily usable tools for the standard-conformant annotation of all the data. The introduction of ISA-Tab is expected to be a first, but important step towards the annotation of biological experiments. Today it's required to make the experimental data on which a publication is based publicly available. In future it should also be required that these data obey the standards and requirements as defined by the MIBBI project and that the submitted data are semantically annotated according to the requirements of the emerging semantic web. Only when biological knowledge is represented in an accurate and comprehensive way, it will be possible for science to effectively make use of these data and for the pharmaceutical and biotechnology industry to develop new efficacious medicines with as few side effects as possible. Only then the trends towards more drought pipelines and decreasing annually numbers of NME's (New Molecular Entities) could be stopped and a drowning in the data pool [2] could be avoided. The same is true for the enhancement of industrial applications of prokaryotic strains and for agricultural and livestock genomics applications. Therefore it's expected that the new occupation of biocuration will arise [481], which should help to address these tasks to be solved efficiently in future. Together with databases and computational services the semantic web is an important part of the envisioned future cyber infrastructure [482,483] for biological sciences. The semantic web und their underlying ontologies will be discussed in a following paper.

## Key Points

- Standards in the biology domain are collected by the central MIBBI project
- The main data management approaches are the use of spreadsheets, web-based document sharing tools, LIS / LIMS / ELN and web service based workflow systems
- If a MISP (Minimum Information of a Software Program) and a standard repository for such a MISP-concordant software are established then workflow systems like Taverna have the potential to become a killer application for the life science field.
- The general options in the selection of a data management system are the development of a proprietary solution, the use of open source and the buying of commercial systems.
- Data integration can be done by link integration, view integration, data warehouses, Service Oriented Architectures, loosely coupled systems, agent based systems and / or visualisation approaches.
- Globally unique identifiers like LSID's are of prime importance for data integration approaches
- There exist a variety of microarray, protein, model and pathway databases into which data can be submitted via TAB-based formats like ISA-TAB.
- To ensure permanent open access to research data publicly funded data centres should be funded to ensure transparency and reuse of data. For this purpose standards and policies for research data deposition and publication need to be defined.

## Acknowledgements

This work was supported by the Federal Ministry of Education and Research (BMBF, Berlin, Germany, HepatoSys systems biology funding initiative, grant 0313080F).